\documentclass[3p,times]{elsarticle}

\usepackage{ecrc}

\volume{00}

\firstpage{1}

\journalname{Nuclear Physics A}

\runauth{R.~Rapp et al.}

\jid{nupha}

\jnltitlelogo{Nuclear Physics A}

\usepackage{graphicx}
\usepackage{color}
\usepackage{amsmath,amssymb}

\newcommand{\erw}[1]{\ensuremath {\left \langle {#1} \right \rangle}}
\newcommand{\dd}{\ensuremath{\mathrm{d}}}
\DeclareMathOperator{\im}{Im}

\begin{document}

\begin{frontmatter}



\title{Properties of Thermal Photons at RHIC and LHC}


\author[label1]{R.~Rapp}
\author[label2]{H.~van Hees}
\author[label3]{M.~He}

\address[label1]{Cyclotron Institute and Department of Physics and Astronomy, Texas A\&M University, College Station, TX 77843, USA}
\address[label2]{FIAS, Ruth-Moufang Str. 1 and Institute for Theoretical Physics, Max-von-Laue-Str. 1, D-60438 Frankfurt, Germany}
\address[label3]{Department of Applied Physics, Nanjing University of Science and Technology, Nanjing 210094, China}

\begin{abstract}
  We study the emission characteristics of thermal photons at RHIC and
  LHC as affected by both the space-time evolution of the bulk medium
  and the thermal emission rates.  For the former we compare the results
  of two evolution models (expanding fireball and hydrodynamics). For
  the latter, we detail the influence of hadronic emission components
  and study a speculative scenario by upscaling the default QGP and
  hadronic rates around the pseudo-critical region.
\end{abstract}


\end{frontmatter}


\section{Introduction}
\label{intro}

The recent observations of a large enhancement of direct photons in
heavy-ion collisions at RHIC~\cite{Adare:2008fq,Adare:2014fwh} and
LHC~\cite{Wilde:2012wc} are indicative for thermal radiation off an
interacting fireball of QCD matter formed in these reactions. It
furthermore turns out that these direct photons carry a 
remarkable elliptic flow ($v_2$)~\cite{Adare:2011zr,Lohner:2012ct},
suggestive for prevalent emission from later stages in the
fireball evolution, when most of the $v_2$ of the bulk has already built
up. Typically, this takes about 5~fm/$c$ by which time the system (in
semi-central Au-Au collisions at RHIC) has cooled down to near the
(pseudo-) critical temperature, $T_{\rm pc}\simeq 170$\,MeV,
cf.~Fig.~\ref{fig1} left. Assuming an average transverse-flow velocity
of $\bar\beta$$\simeq$\,0.4$c$ around this region yields a schematic
estimate of the inverse-slope parameter of such a source of $T_{\rm
  eff}\simeq T_{\rm pc}
\sqrt{(1+\erw{v_T})/(1-\erw{v_T})}\simeq$\,255\,MeV, which is within the
range of values measured by PHENIX~\cite{Adare:2008fq,Adare:2014fwh},
see Fig.~\ref{fig1}~\cite{vanHees:2011vb}. Such a source is compatible
with what has been deduced from other observables, e.g., for the
quark distribution functions used in phenomenological coalescence models
to reproduce the constituent-quark scaling of light and multi-strange
hadron spectra and $v_2$, or for thermal dileptons, where large emission
contributions from around $T_{\rm pc}$~\cite{Rapp:2009yu,Rapp:2013nxa}
explain the apparent dissolution of the $\rho$ resonance from SPS to
RHIC energies. Especially the latter is closely related to photons which
correspond to the $M\to0$ limit of dileptons. This link has been
explored in Ref.~\cite{vanHees:2011vb}; using a schematic fireball model
for the bulk evolution with state-of-the-art photon rates, the
direct-photon $v_2$ turns out to be not far from the PHENIX data. In
particular the hadronic rates are larger than in other calculations in
the literature. Most notably, the $M\to0$ limit of the $\rho$ spectral
function~\cite{Turbide:2003si} includes a rather extensive set of
baryon-induced processes, known to be of critical importance for the
low-mass dilepton enhancement~\cite{Rapp:2009yu}. In the following, we
scrutinize these findings by testing various ingredients of the thermal
photon calculations, both with respect to the space-time evolution
(utilizing hydrodynamics with initial flow and sequential chemical and
kinetic freezeout) and the photon rates (switching off the $\rho$
spectral function contribution, and upscaling current QGP and hadronic
rates around $T_{\rm pc}$)~\cite{vanHees:2014ida}.


\section{Thermal Emission Rates}
\label{sec_rates}
In thermal field theory the equilibrium emission rate of photons,
\begin{equation}
\label{rate}
q_0\frac{\dd N_{\gamma}}{\dd^4 x \dd^3 q} = -\frac{\alpha_\mathrm{EM}}{\pi^2}
\ f^{\mathrm{B}}(q_0;T)  \im \Pi_\mathrm{EM}^T(q_0=q;\mu_B,T), 
\end{equation}
is determined by the in-medium photon selfenergy,
$\Pi_\mathrm{EM}^T$, due to the coupling to the partonic or
hadronic constituents of the thermal bath. Its imaginary part
represents the different cuts of the corresponding Feynman
diagrams which characterize the scattering processes for thermal
photon production.

\begin{figure}
\begin{center}
\includegraphics*[width=7.2cm]{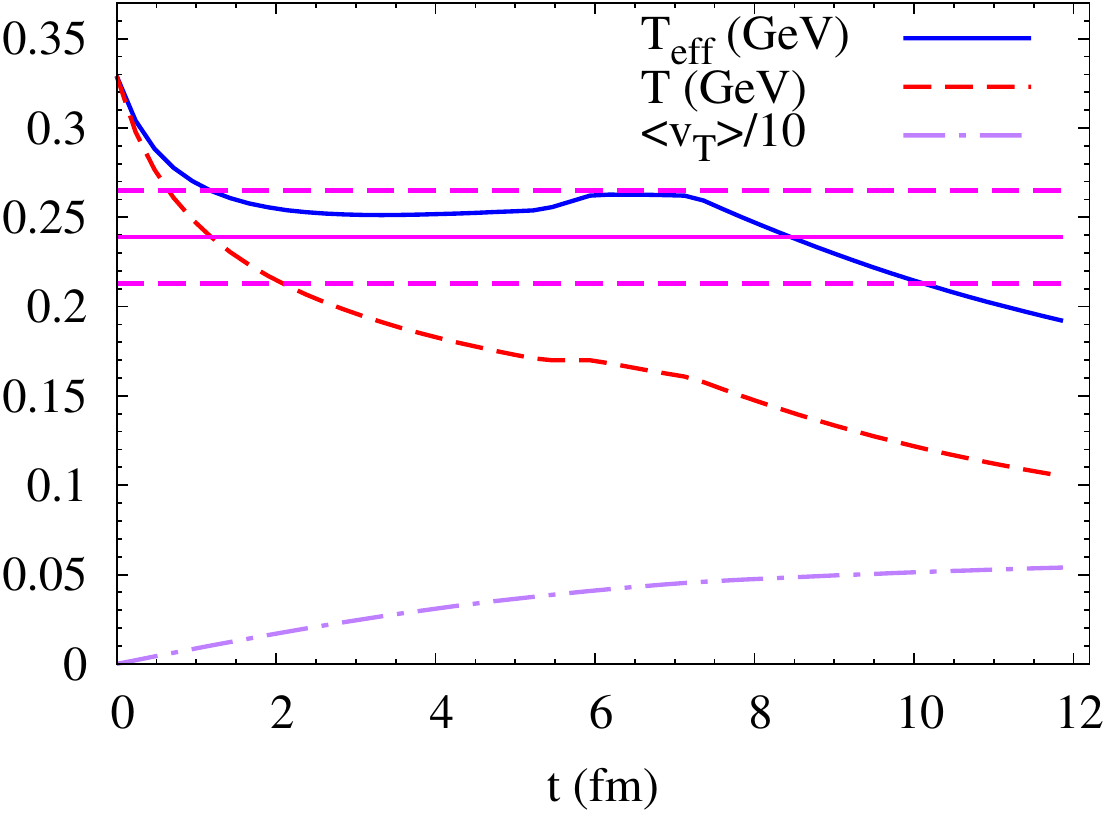}
\hspace{0.5cm}
\includegraphics*[width=7.5cm]{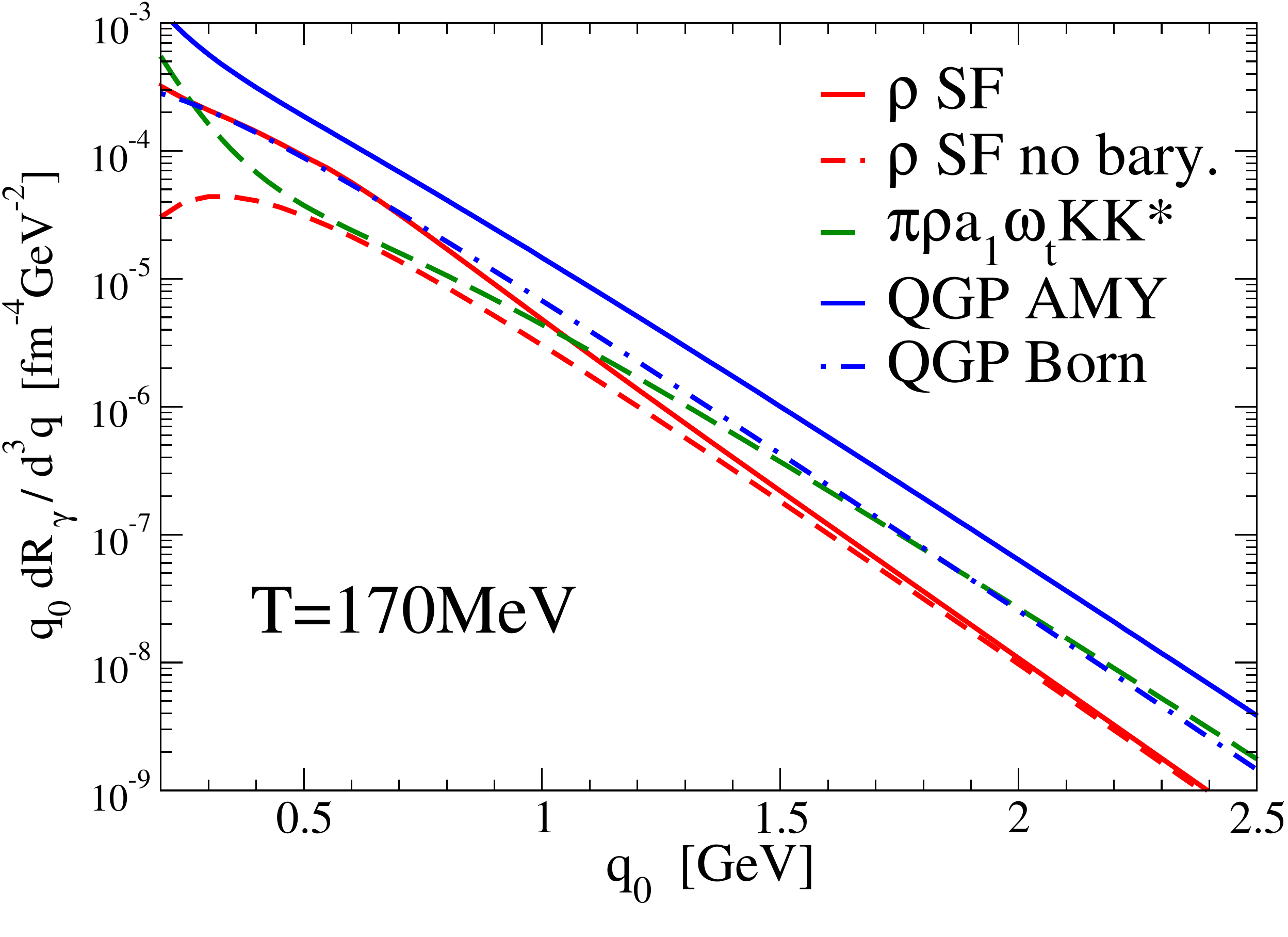}
\caption{Left panel: time dependence of inverse slope
  parameter of the photon $q_t$ spectrum (solid blue line), rest frame
  temperature (red dashed line) and average transverse expansion
  velocity (dash-dotted line) from a fireball evolution, compared to the
  measured inverse slope of the excess radiation in 0-20\%
  Au-Au($\sqrt{s_{NN}}$=200\,GeV), $T_{\rm
    slope}$=(239$\pm$26)\,MeV~\cite{Adare:2014fwh} (pink horizontal
  lines).  Right panel: thermal photon production rates from the QGP
  (solid blue line: full leading order~\cite{Arnold:2001ms}, dash-dotted
  line: Born rate~\cite{Kapusta:1991qp}) and hadronic matter (red solid
  and dashed line: $\rho$ spectral function contribution with and
  without baryons; long-dashed line: tree-level meson gas
  reactions~\cite{Turbide:2003si}).}
\label{fig1}
\end{center}
\end{figure}
For photon emission from the QGP we employ the complete
leading-order results of Ref.~\cite{Arnold:2001ms}. For emission
from hadronic matter we employ the zero-mass limit of the $\rho$
spectral function of Ref.~\cite{Rapp:1999us} which includes baryon and
meson resonances resulting in pertinent Dalitz decays, as well as
pion-exchange reactions with baryons generated by the in-medium pion
cloud; in addition, pion, kaon, $a_1$ and $\omega$ $t$-channel exchange
reactions in a $\pi\rho KK^*$ gas (not included in the $\rho$ spectral
function) have been added through pertinent Born diagrams using the
standard kinetic theory expression~\cite{Turbide:2003si}. These rates
are summarized for a temperature of $T$=170\,MeV and vanishing baryon
chemical potential ($\mu_\mathrm{B}$=0) in Fig.~\ref{fig1} right. 
Contributions from anti-/baryons in hadronic matter are significant up 
to rest-frame momenta of $q_0$$\le$1\,GeV; for typical hadronic flow 
velocities at RHIC and LHC, this translates into lab-frame momenta of 
up to $\sim$2\,GeV. Meson gas $t$-channel exchange reactions take over for
$q_0$$>$1\,GeV. The sum of all hadronic processes is comparable to the
perturbative QGP rates.

\section{Space-Time Evolution}
\label{sec_evo}

To model the space-time evolution of the thermal medium, over which the
rates have to be integrated, we have considered two approaches: 
(i) an expanding fireball with ellipsoidal transverse area,
assumed to be isotropic at each time step and with conserved total
  entropy~\cite{vanHees:2011vb}; (ii) a (2+1)-D ideal hydrodynamic
model~\cite{He:2011zx} based on the original code of
Ref.~\cite{Kolb:2003dz}.  Both models utilize the same equation of state
(lattice EoS for the QGP with a near-smooth transition to a hadron
resonance gas at $T_{\rm pc}$=170\,MeV, followed by hadro-chemical
freezeout at $T_{\rm ch}$=160MeV), and have been adjusted to the same
set of light and multistrange hadron spectra and $v_2$. Kinetic
freezeout for multistrange hadrons has been imposed at $T_{\rm ch}$,
facilitated by a suitable transverse acceleration in the fireball model, 
and by a non-vanishing initial flow in the hydro
model. For both models this implies that the bulk-medium $v_2$ levels
off close to $T_{\rm ch}$, which, in turn, also helps in a better
description of the pion and proton $v_2$ at kinetic freezeout at
$T_{\rm kin}$$\simeq$\,110\,MeV.

\section{Photon Spectra and Flow}
\label{sec_spec}
\begin{figure}
\begin{center}
\includegraphics*[width=7.9cm]{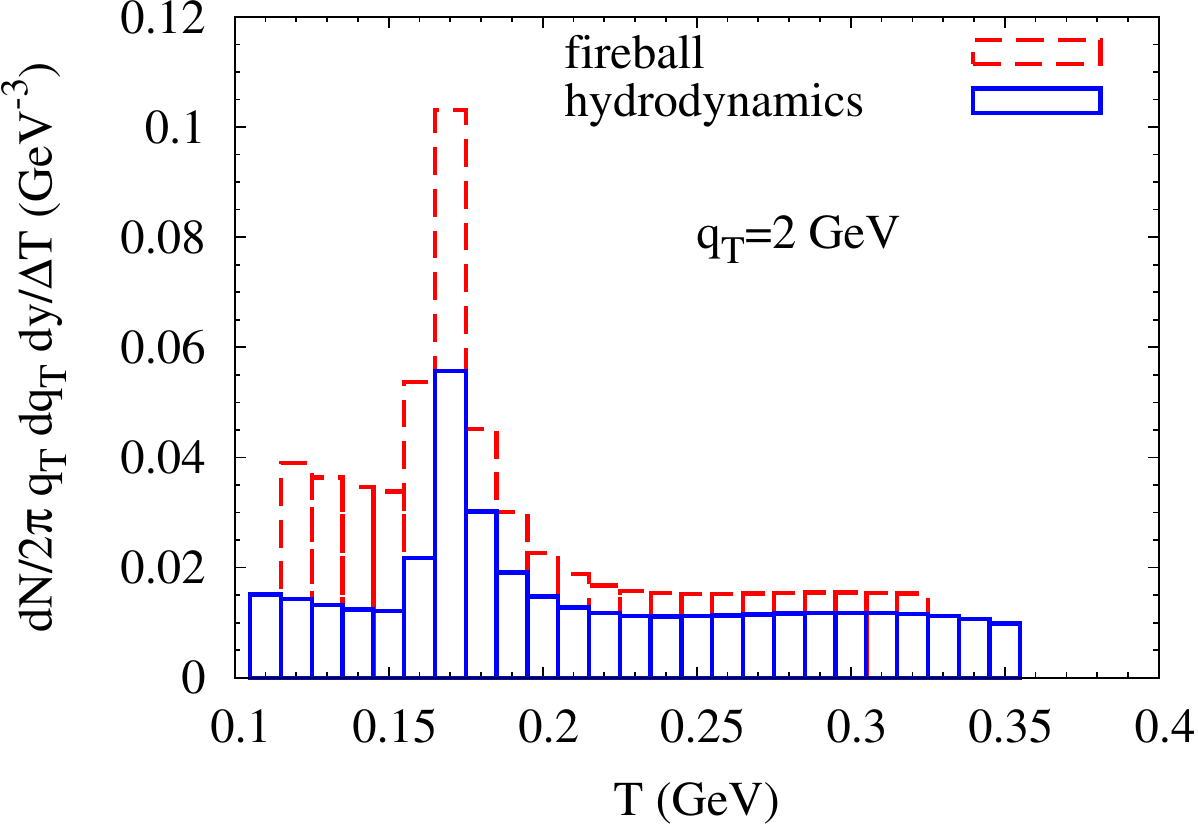}
\hspace{0.3cm}
\includegraphics*[width=7.9cm]{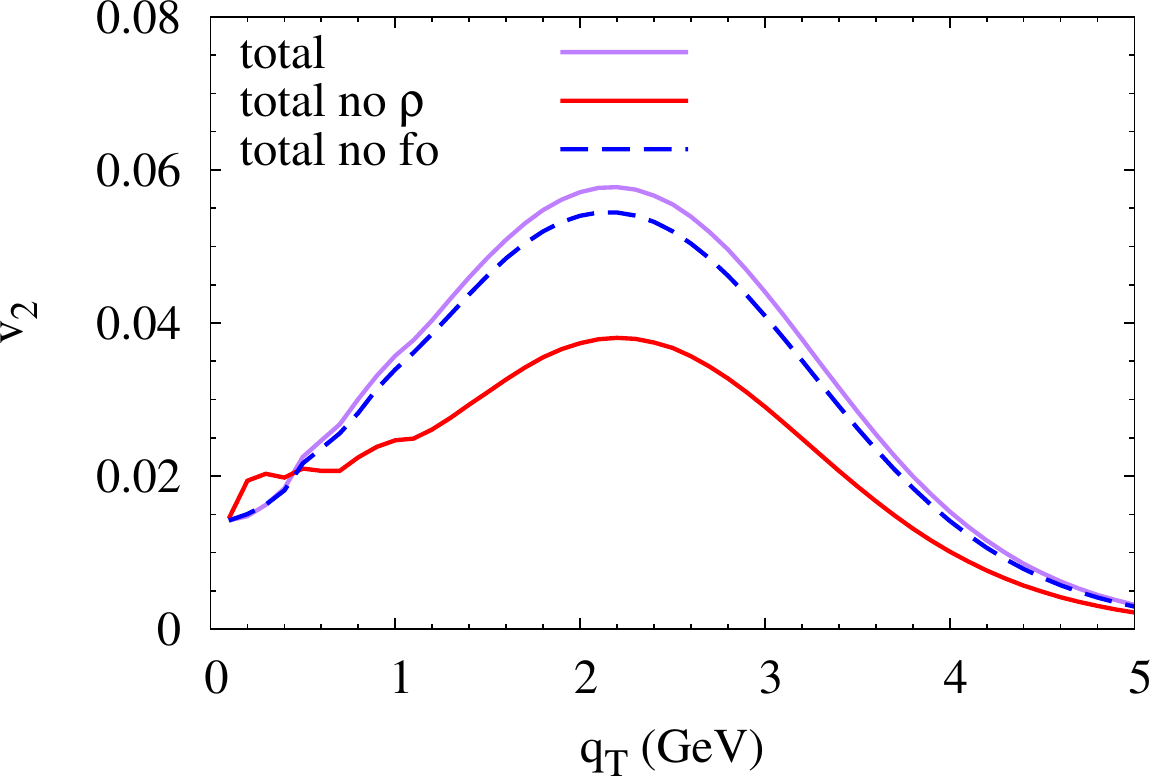}
\caption{Left: temperature emission profile of thermal photon yields
  from fireball and hydro evolutions at $q_T$=2\,GeV. Right: reduction 
  of the total direct photon $v_2$ from the fireball when switching off
  the contributions from either the in-medium $\rho$ spectral function or
  strong final-state decays.}
\label{fig:contr}
\end{center}
\end{figure}
A detailed comparison of the photon spectra from the fireball and hydro
evolution (using the same default rates, as plotted in Fig.~\ref{fig1})
reveals a reasonable agreement (within ca.~20\%) of the QGP spectra;
both space-time models also exhibit a pronounced maximum of the temperature
differential emission around $T_{\rm pc}$ up to momenta of at least
$q_T$=2\,GeV, see Fig.~\ref{fig:contr} left. This is a consequence of
the remnant of the latent heat in the transition region, i.e., the large
variation in entropy density over a small range in $T$ (similar to
low-mass dileptons). However, in the hadronic phase, the hydro emission
falls significantly below the one from the fireball, more so at higher
momenta (up to ca.~50\% at $q_T$=2GeV in Au-Au at RHIC). Part of the
reason is the continuous freeze-out of the hydro medium, while the
entire fireball volume only freezes out at the end of the
evolution. While the latter is clearly an over-simplification, one
should  note that hadrons which are frozen out on the hydro
hypersurface never re-thermalize, although, in principle, they could be
reabsorbed by the freeze-out front. Therefore, we expect the most
realistic hadronic emission in between the hydro and fireball results.
We believe that another part of the discrepancy in the hadronic photon 
emission spectra, which becomes more pronounced toward higher $q_T$, 
is the ``inward'' burning of the freeze-out front. This leads to rather
low-flow hadronic freeze-out cells especially toward the end of the
evolution. On the contrary, the outward moving fireball front implies a
monotonously increasing average flow (cf.~Fig.~\ref{fig1} left). The
``inward burning'' of the hydro was identified as a possible problem in
reproducing the measured HBT radii of pions. The initial flow introduced
into our hydro evolution helps with this, but possibly not enough.

\begin{figure}
\begin{center}
\includegraphics*[width=8cm]{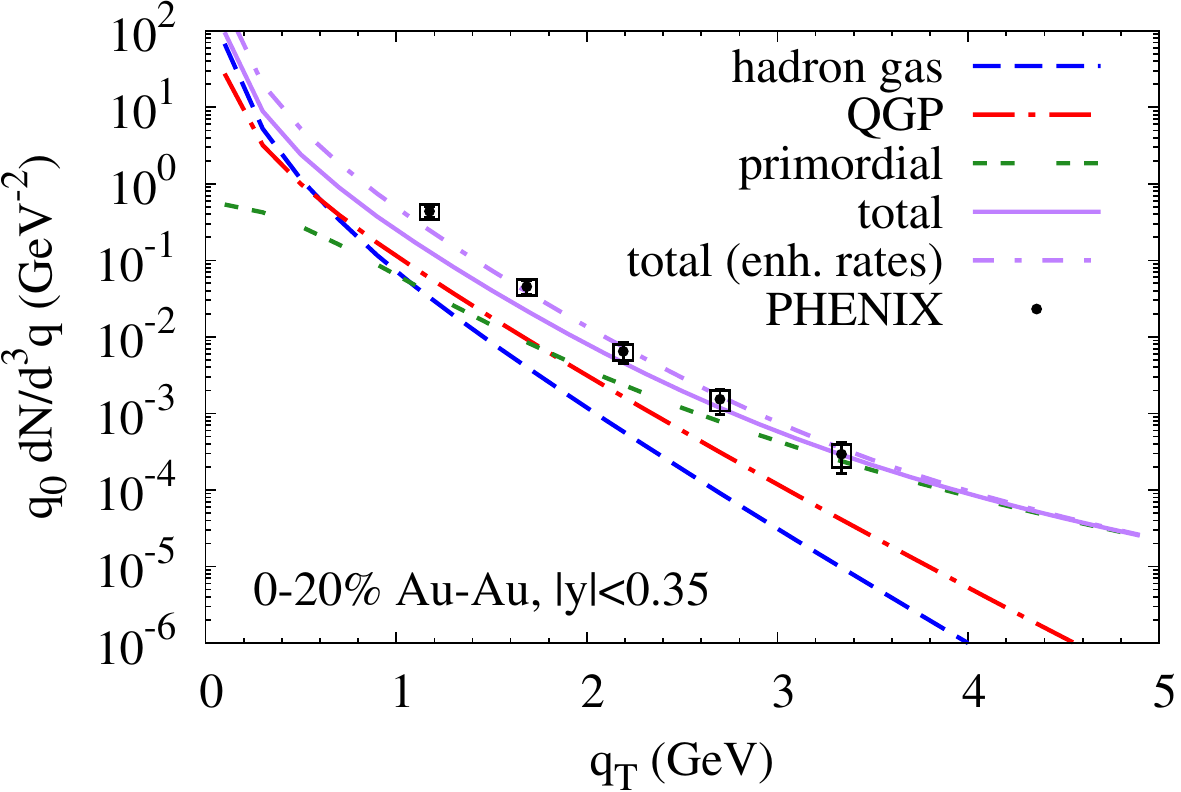}
\hspace{0.3cm}
\includegraphics*[width=7.9cm]{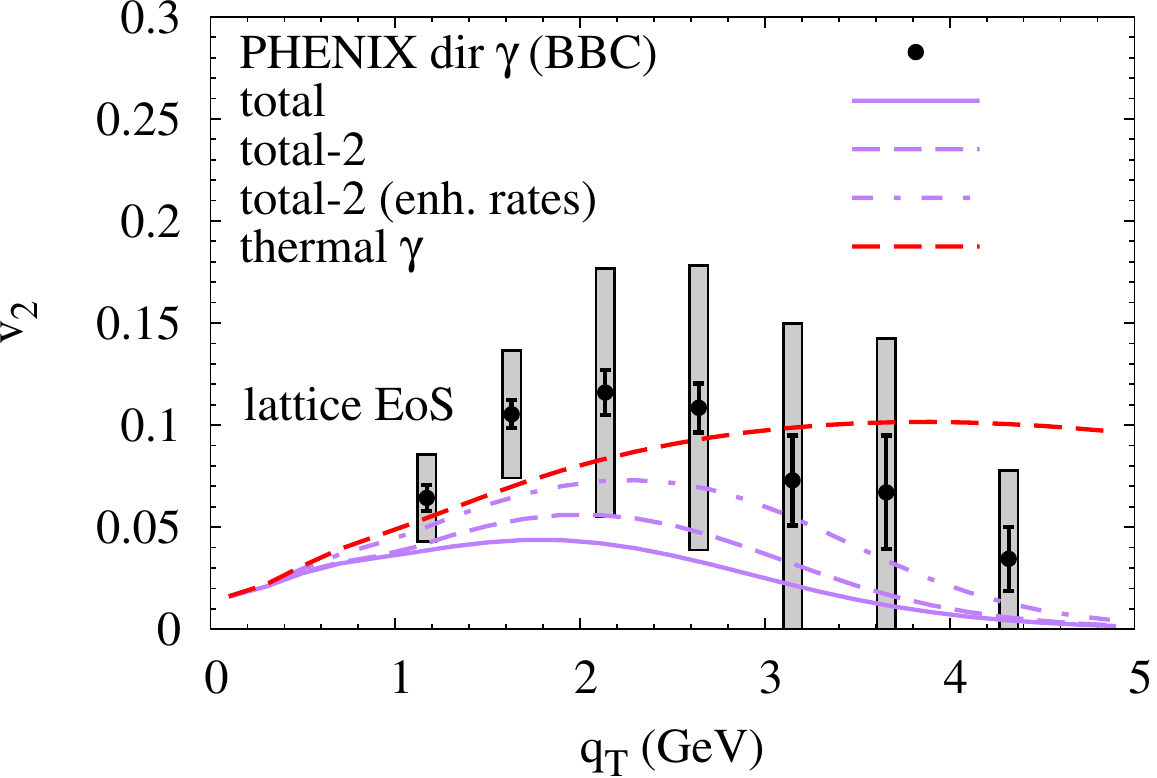}
\caption{Direct photon spectra (left) and elliptic flow (right) from the
  hydrodynamic evolution using the default emission rates with the
  PHENIX pp reference for primordial emission (solid lines), with a
  pQCD-motivated primordial emission (dashed line in the right panel),
  and with ``pseudo-critically'' enhanced rates (see text for details)
  with pQCD primordial (dash-dotted lines). PHENIX data are from
  Refs.~\cite{Adare:2008fq,Adare:2011zr}.}
\label{fig:ph-obs}
\end{center}
\end{figure}

As a result of the discrepancy in the hadronic emission, the total
direct-photon spectra from the hydro evolution (plus primordial photons)
in 0-20\% Au-Au at RHIC are below the fireball results by a few tens of
percent in the region where thermal radiation is prevalent. Likewise,
the maximum in the photon $v_2$ is about 25\% smaller for the hydro
evolution. Yet our hydro results for the photon $v_2$ are not
far from the lower limit of the error bars of the PHENIX data,
especially when allowing for moderate modifications of the primordial
contribution, see Fig.~\ref{fig:ph-obs} right. This constitutes an
improvement over existing calculations. We attribute the
main differences to the larger hadronic emission rates, the initial flow
in the hydro evolution, and the strong-decay feeddown (e.g., $\Delta\to
N\gamma$, $a_1\to \pi\gamma$, etc.) in our calculation. For example,
when switching off the (hadronic) photon contributions from the
in-medium $\rho$ spectral function, the maximum total direct-photon
$v_2$ in the fireball calculation decreases by about 35\%, while
switching off the strong feeddown reduces it by nearly 10\%,
cf.~Fig.~\ref{fig:contr} right.  At the LHC, the stronger QGP dominance
renders closer agreement between hydro and fireball results. Both 
calculations tend to underestimate the preliminary ALICE $q_T$ spectra,
but agree with the preliminary $v_2$ data within experimental errors.

Since the default thermal emission rates do not fully account for the
observed direct-photon enhancement, we investigated an extreme scenario
of upscaling our rates by a factor reaching a maximum of 3 around
$T_{\rm pc}$ (``pseudo-critical enhancement''), and convoluted these
rates over the hydro medium. This basically resolves the discrepancies
with both RHIC and LHC spectra and $v_2$, see, e.g., the dash-dotted
lines in Fig.~\ref{fig:ph-obs}. However, it is presently unclear whether
such an enhancement can be substantiated by microscopic rate calculations, 
and, if confirmed, what ramifications this would have for dilepton 
observables (where current theory and data agree~\cite{Rapp:2013nxa}).

\section{Conclusions}
\label{sec_concl}

We have computed thermal photon spectra and $v_2$ in heavy-ion
collisions at RHIC and LHC, using state-of-the-art thermal emission
rates from QGP and hadronic matter including the effect of baryons which
importantly figure in the dilepton context. We have compared two
different evolution models based on the same EoS and fits to bulk-hadron
data, as well as the concept of sequential freezeout. Their main
difference appears in the hadronic emission spectra, which are larger in
the thermal fireball than in the hydro evolution. Consequently, the
fireball does slightly better in describing the experimental excess
spectra and $v_2$ of direct photons at RHIC, even though the hydro
results are not far from the data either given their current
uncertainties. At LHC, both models tend to underestimate the 
spectra but are compatible with the $v_2$ from preliminary ALICE 
measurements. Important ingredients in our approach are
large hadronic rates and a rapid built-up of radial flow (necessary to
implement sequential freeze-out). An upscaling of the thermal rates
around $T_{\rm pc}$ further improves the agreement, but a microscopic
calculation of such an effect remains an open issue.

{\bf Acknowledgemts}. This work has been supported by the grants U.S.-NSF 
PHY-1306359, NSFC 11305089, BMBF 05P12RFFTS, and by LOEWE through HIC 
for FAIR.






\begin{thebibliography}{00}

\bibitem{Adare:2008fq}
  A.~Adare {\it et al.}  [PHENIX Collaboration],
  Phys.\ Rev.\ Lett.\  {\bf 104} (2010) 132301.

\bibitem{Adare:2014fwh}
 A.~Adare {\it et al.}  [PHENIX Collaboration],
  arXiv:1405.3940 [nucl-ex].

\bibitem{Wilde:2012wc}
  M.~Wilde [ALICE Collaboration],
  Nucl.\ Phys.\ A {\bf 904-905}, 573c (2013).

\bibitem{Adare:2011zr}
  A.~Adare {\it et al.}  [PHENIX Collaboration],
  Phys.\ Rev.\ Lett.\  {\bf 109}, 122302 (2012).

\bibitem{Lohner:2012ct}
  D.~Lohner {\it et al.} [ALICE Collaboration],
  J.\ Phys.\ Conf.\ Ser.\  {\bf 446}, 012028 (2013).


\bibitem{vanHees:2011vb}
  H.~van Hees, C.~Gale and R.~Rapp,
  Phys.\ Rev.\ C {\bf 84}, 054906 (2011).

\bibitem{Rapp:2009yu}
  R.~Rapp, J.~Wambach and H.~van Hees,
  in \emph{Relativistic Heavy-Ion Physics}, edited by R.~Stock and
  Landolt B\"ornstein (Springer), New Series {\bf I/23A}, 4-1 (2010);
  [{\tt arXiv:0901.3289[hep-ph]}].

\bibitem{Rapp:2013nxa}
  R.~Rapp,
  Adv.\ High Energy Phys.\  {\bf 2013} (2013) 148253.

\bibitem{Turbide:2003si}
  S.~Turbide, R.~Rapp and C.~Gale,
  Phys.\ Rev.\  C {\bf 69}, 014903 (2004).

\bibitem{vanHees:2014ida}
  H.~van Hees, M.~He and R.~Rapp,
  submitted to Nucl. Phys. {\bf A} (2014); arXiv:1404.2846 [nucl-th].

\bibitem{Arnold:2001ms}
P.B.~Arnold, G.D.~Moore and L.G.~Yaffe,
  JHEP {\bf 0112}, 009 (2001).

\bibitem{Rapp:1999us}
  R.~Rapp and J.~Wambach,
  Eur.\ Phys.\ J.\ A {\bf 6} (1999) 415.

\bibitem{Kapusta:1991qp}
  J.I.~Kapusta, P.~Lichard and D.~Seibert,
  Phys.\ Rev.\ D {\bf 44} (1991) 2774
   [Erratum-ibid.\ D {\bf 47} (1993) 4171].

\bibitem{Kolb:2003dz}
  P.F.~Kolb and U.W.~Heinz,
  In R.~C.~Hwa (ed.) et al.: Quark gluon plasma, 634, [nucl-th/0305084].

\bibitem{He:2011zx}
  M.~He, R.J.~Fries and R.~Rapp,
  Phys.\ Rev.\ C {\bf 85}, 044911 (2012).





\end{thebibliography}



\end{document}